\documentclass[10pt]{article}
\title{Relativistic Rotation in the Large Radius,\\Small Angular Velocity Limit}
\author{Robert D. Klauber\\1100 University Manor Dr., 38B, Fairfield, IA 52556, USA\\email: rklauber@netscape.net}
\date{September 7, 2002}
\usepackage {graphicx}
\usepackage {longtable}
\usepackage{hyperref}
\begin{document}
%\draft
\maketitle
\begin{abstract}
% put the following at beginning of ref section just after begin{references} command
% for RevTex I believe
%\bibitem[*]{email}rklauber@netscape.net.

%\pacs{04.20,03.30+p}

Relativistic rotation is considered in the limit of angular velocity 
approaching zero and radial distance approaching infinity, such that 
centrifugal acceleration is immeasurably small while tangent velocity 
remains close to the speed of light. For this case, the predictions of the 
traditional approach to relativistic rotation using local co-moving Lorentz 
frames are compared and contrasted with those of the differential geometry 
based non-time-orthogonal analysis approach. Different predictions by the 
two approaches imply that only the non-time-orthogonal approach is valid.

\end{abstract}

\section{Introduction}
\label{sec:introduction}

An approach to analyzing relativistic rotation that has been popular since 
the early 1900s entails use of local Lorentz frames located throughout a 
rotating 
frame\cite{Relativistic:1977}$^{,}$\cite{Einstein:1938}$^{,}$\cite{Stachel:1980}$^{,}$\cite{Weber:1997}$^{,}$\cite{Pathra:1}$^{,}$\cite{Richard:1944}. 
Each local Lorentz frame has an instantaneous velocity with respect to the center 
of rotation equal to the rotating frame tangent velocity ${\rm {\bf v}} = 
\mbox{\boldmath $\omega$} \;\times \;{\rm {\bf r}}$ at the point within the rotating frame 
where the Lorentz frame is located. The desired analysis is done locally 
using special relativity and then integrated over successive local Lorentz 
frames to obtain global values which one would presumably measure in 
experiment. We shall refer to this method as the ``traditional approach'' to 
relativistic rotation.

Proponents of this approach generally consider the rim of a rotating disk to 
undergo Lorentz contraction\cite{Actually:1}$^{,}$\cite{Arzeli:1966}. This 
conclusion is based on the well-known velocity dependent Lorentz contraction 
that occurs in each local frame, followed by a spatial integration from 
local frame to local frame all around the disk rim.

Proponents of the traditional approach often cite, as support for the 
approach, the case of very large radius $r$, vanishingly small $\omega $, where 
$v=\omega r$ is on the order of $c$, the speed of light, and centrifugal 
acceleration $a = \omega ^{2}r$ is too small to be measured in any experiment. 
We refer to this case herein as the ``limit case''. In such case, it is 
reasoned, the local region at radius $r$ can not be discerned from a true 
Lorentz frame, since neither $\omega $ nor $\omega ^{2}r$ are different 
enough from zero to be detectable. Thus, an observer on the rim of such a 
large, slowly rotating disk, and a second observer adjacent the first but 
stationary in the non-rotating ``lab'', can be considered fixed in separate 
Lorentz frames. It follows that traditional special relativity theory may 
then be applied directly in all regards.

An alternative method for analysis of rotation comprises use of differential 
geometry\cite{Paul:1937}$^{,}$\cite{Robert:1998} and suitable 
transformations between the rotating and lab frames. As this leads to a 
rotating frame metric with off diagonal terms, and hence a non-orthogonality 
between space and time, we refer to this method as the non-time-orthogonal 
(NTO) approach. The predictions of the traditional and NTO approaches agree 
in many, but not quite all, 
respects\cite{Neil:1997}$^{,}$\cite{Brillet:1979}$^{,}$\cite{Sagnac:1914}$^{,}$\cite{Alexandre:1942}$^{,}$\cite{Robert:1}$^{,}$\cite{Ref:1}.

In this article we compare and contrast the predictions of the traditional 
and NTO approaches specifically in the limit case, and primarily with regard 
to Lorentz contraction. Prior to doing so, we make several observations for 
the non-limit case where acceleration and angular velocity are readily 
measurable.

\section{The Traditional Approach}
\label{sec:mylabel1}

Consider a rotating disk first in the non-limit case wherein $\omega $ and 
$\omega ^{2}r$ are significant and can be readily measured. According to 
the traditional approach, an observer in the lab sees rods aligned with, and 
fixed to, the rim of the disk Lorentz contract. It is then concluded that 
measurement of the circumference $C$ and the radius $r$ will result in $C \ne 2\pi 
r$ and therefore the 2D surface of the rotating disk must be curved.

There are two problems with this conclusion. First, according to special 
relativity, an observer in one frame sees rods in the frame of a second 
observer (with relative velocity difference from the first) as contracted, 
but the second observer sees no such contraction of her own rods. Thus, the 
lab observer may see the disk rim rods as contracted, but an observer on the 
rim would notice no such contraction. If the disk observer sees no 
contraction, then the Riemann curvature she measures for her disk surface 
must be zero, and the surface must be flat\cite{Tartaglia:1999}. 

Second, again according to special relativity, each observer sees the 
other's rods as contracted. Hence, by the traditional logic, the disk 
observer must see the lab rods as contracted, and therefore conclude that 
the lab surface is curved. But those of us living in the lab know that this 
is simply not true, and the analysis appears inconsistent.

In response to these points, I have heard the argument that special 
relativity does not apply to the disk observer, as her frame is not 
inertial. My response to this is several fold.

First, I ask for mathematical analysis (real physics, not mere words) 
showing how Lorentz contraction then occurs on the disk rim in an absolute 
way, i.e., so that both the disk and lab observers agree that the disk meter 
sticks are shorter than the lab meter sticks. No one has yet provided this.

Second, the fundamental assumption in using local Lorentz frames is that 
locally a non-inertial (curved) frame can be represented by an inertial 
(flat) frame. How can we assume local inertial frames can be used as 
surrogates at the beginning of the analysis to draw conclusions, and then 
when inconsistencies arise at the end, simply say that they can 
not\cite{NTO:1}?

Third, in the traditional analysis limit case, the disk observer is 
considered effectively inertial and all of special relativity applies 
directly. Thus, she should not see her own meter sticks as contracted, 
thereby meaning that her disk surface is flat. Yet, nothing in passing to 
the limit case implied anything other than $C$ remaining $ \ne 2\pi r$ in that 
limit.

Fourth, in the limit case the lab meter sticks have to appear contracted to 
the disk observer, since said observer is now Lorentzian. This means, 
according to the traditional approach logic, that the lab surface is curved. 
There is no ``out'' in this case of claiming the disk is not really 
inertial, such that things are somehow different from standard special 
relativity.

Finally, if the contraction for rotation is argued to be absolute in the 
non-limit case, then it must also be so in the limit case. Yet in the limit 
case, both observers are considered to be Lorentzian, and hence the Lorentz 
contraction must be relative (each must see the other's rods as contracted) 
and can not be absolute.

The traditional analysis thus appears to possess serious internal 
contradictions.

\section{Non-time-orthogonal Approach}

\subsection{Background}

As noted earlier, analysis of rotating frames using differential geometry 
and the most widely accepted transformation between the lab and the rotating 
frame leads to a rotating frame metric with off diagonal time-space 
components. Thus, in such a frame time is not orthogonal to space (NTO).

As noted in endnote \cite{Neil:1997} the NTO (or Langevin) metric 
correctly predicts GPS measured results for the rotating earth frame, 
whereas the traditional approach does not. Further, as noted in ref. 
\cite{Robert:1998}, while the NTO approach predicts the traditional time 
dilation and mass energy increase measured in many cyclotron (i.e., 
rotation) experiments, it predicts no Lorentz contraction in rotating 
frames. 

\subsection{NTO Non-limit Case}

If, as predicted by NTO theory, there is no Lorentz contraction for rotating 
frames, in either an absolute or relative sense, then the non-limit case 
objections to the traditional approach of Section \ref{sec:mylabel1} 
vanish.

In NTO theory, by testing one can distinguish between a velocity associated 
with a rotating frame (e.g., the velocity of a point on the rim of a 
rotating disk) and a velocity associated with an inertial frame. A Foucault 
pendulum, or any number of means, detects angular velocity $\omega $. A 
spring mass system measures $\omega ^{2}r$, and from it the distance $r$ to 
the center of rotation can be determined. The tangential velocity of any 
point within the rotating frame is thus $v = \omega r$, and this can be 
determined in an absolute sense from within the rotating frame without 
looking outside. No such absolute determination of velocity is possible for 
a translating frame.  (Note that for simplicity we illustrate this and related 
points herein using non-relativistic mechanics.  Analogous relativistic 
analyses lead to the same conclusions.\cite{For:1})

This ability to distinguish between velocities arising from rotation and 
translation lies at the foundation of NTO theory (see ref. 
\cite{Robert:1998}.) In fact, the degree of non-time-orthogonality (the 
slope of the time axis with respect to the circumferential space direction 
axis) is directly related to $v = \omega r$.

\section{Challenges to the NTO Approach }
\label{sec:challenges}

I have been challenged in private communications to defend NTO analysis in 
the following ways.

1) Consider a meter stick bolted to, and aligned with the circumference of, a 
rotating disk. An observer fixed in the lab adjacent the moving rim sees no 
Lorentz contraction of the disk meter stick (according to NTO theory.) 
Suddenly the bolt breaks and the meter stick is released. As the meter stick 
then instantaneously changes from an NTO to a Lorentz frame, it should 
appear to the lab observer to contract instantaneously. This, it is argued, 
is unnatural and even impossible. There is no change in velocity, yet there 
is significant change in meter stick length.

2)  Consider the limit case of infinitesimally small $\omega $ and 
infinitesimally small $\omega ^{2}r$. Via NTO analysis there is no Lorentz 
contraction even though $v = \omega r$ is, say, $c$/2. If there is no possible means 
to detect angular velocity or centrifugal acceleration, then how can one 
distinguish between a local Lorentz frame with speed $c$/2 relative to the lab 
and the presumed NTO frame of the disk? That is, how can a meter stick in 
the former frame look contracted, whereas one in the latter does not, if one 
can measure no other difference between the two frames.

3)  Consider combining 1) and 2) 
above. In the limit case, the meter stick attached to the disk is suddenly 
released. To the lab observer it instantaneously contracts. Apart from the 
instantaneous nature of the contraction [as already questioned in 1)], there 
is no way to distinguish between the pre and post release states of the 
meter stick apart from the contraction. That is, for no detectable change in 
$\omega $, $\omega ^{2}r$, or apparently any other quantity, there is a 
very significant change in length. Stated somewhat differently, for a nearly 
infinitesimally small change in $\omega $ and $\omega ^{2}r$ (from 
immeasurably close to zero to zero), there is a finite change in observed 
length.

\section{Replies to the NTO Challenges}
\label{sec:replies}

\subsection{Preliminary Comments}

Of course, given enough time, or a large enough spatial region, one can 
always discern whether one's own frame (or another frame) is rotating or 
not. There are many ways to do this, e.g., a Foucault pendulum, a 
measurement of Coriolis type motion of free particles, a Sagnac experiment, 
etc. In particular, one could compare a meter stick pinned to one's frame 
with a free fall meter stick. The distance between the two, and the angle 
formed between the two, will change over time, albeit extremely slowly in 
the limit case. From these changes one can calculate the value for $v = \omega 
r$, and thence the expected NTO effects and their variation from standard 
relativistic effects. (As shown in reference \cite{Robert:1998} the 
degree of non-time-orthogonality is directly related to $\omega r$.)

Additionally, nature always knows the values for $\omega $ and $\omega 
^{2}r$ (and hence $v = \omega r$.) Even if they are too subtle for us to measure 
over practical time and spatial intervals, the physical world effect should 
still be there, and it should agree with our theoretical calculations (given 
that the theory is correct.) By analogy, consider a radial length in a 
cylindrical coordinate system that is light years in length. A near 
infinitesimally small change in angle of the radial line near the origin 
results instantaneously in a very large displacement circumferentially at 
the other end. Immeasurably small changes in a system can indeed cause other 
finite, and readily measurable changes.

However, proponents of the traditional theory advocate use of local Lorentz 
frames, whose validity for analyzing non-inertial cases has in the past been 
considered beyond repute. Such Lorentz frames are localized in both time and 
space, i.e., they have infinitesimal extent in four dimensions. Hence, it is 
argued, they should be equivalent to the non-inertial frame in question for 
periods of time and space over which rotation can not be discerned.

I argue that locally time is not orthogonal to space and the effective 
metric to use locally (which would yield measured values for time and space, 
i.e., values measured with standard meter sticks and clocks) is

\begin{equation}
\label{eq1}
g_{\mu \nu } = \left[ {{\begin{array}{*{20}c}
 { - 1} \hfill & 0 \hfill & b \hfill & 0 \hfill \\
 0 \hfill & 1 \hfill & 0 \hfill & 0 \hfill \\
 b \hfill & 0 \hfill & 1 \hfill & 0 \hfill \\
 0 \hfill & 0 \hfill & 0 \hfill & 1 \hfill \\
\end{array} }} \right].
\end{equation}

The off diagonal quantity $b$ is a function of $\omega r$. When $\omega $ = 0, 
then $b$ = 0, and the metric reduces to the Lorentz/Minkowski metric $\eta 
_{\mu \nu } = diag( - 1,1,1,1)$. Metric (\ref{eq1}) is the general case local 
metric. The Minkowski metric, representing a time orthogonal Lorentz frame, 
is a special case (albeit by far the most common case.) Globally, we can 
measure $\omega $ readily, just as we can measure average velocity of an 
object than travels at constant velocity. In the latter case, the local 
velocity over infinitesimal time and distance scales (which we can not 
measure) equals the global average velocity (which we can measure). 
Similarly, the globally measured $\omega $ and the concomitant global 
non-time-orthogonality are mirrored locally, though we are unable to measure 
them locally.

Regardless of the relevance one may attribute to the above remarks, I will 
respond to the challenges of section \ref{sec:challenges} with 
arguments that are not dependent on them.

\subsection{Additional NTO Analysis Background}
\label{subsec:additional}

In Section 5.3 of reference \cite{Robert:1998} I note there is another 
way to measure $\omega r$. If one is on the rotating disk, one experiences a 
potential energy that equals the integral of the force felt ($m\omega 
^{2}r)$ over the radial distance from the rim to the center. One feels 
oneself inside an effective gravitational field. In classical theory the 
potential energy is simply $ - \textstyle{1 \over 2}m\omega ^2r^2$. In 
general relativity total energy of a particle at rest in a potential field 
is

\begin{equation}
\label{eq2}
e = mc^2\sqrt {1 + \frac{2\Phi }{c^2}} ,
\end{equation}

\noindent
where $\Phi $ is the classical (Newtonian) potential (potential energy per 
unit mass). Thus one finds\cite{NTO:2}

\begin{equation}
\label{eq3}
\begin{array}{c}
 e = mc^2\sqrt {1 - \textstyle{{\omega ^2r^2} \over {c^2}}} = mc^2 - 
\textstyle{1 \over 2}m\omega ^2r^2 - (higher\;order\;terms) \\ 
 = mc^2 + V_{classical} + \;...... = mc^2 + V_{relativisitc} \\ 
 \end{array}
\end{equation}

\noindent
where $V_{classical}=m\Phi $. The total energy is thus the familiar rest 
energy \textit{mc}$^{2}$ plus the relativistic potential energy.

In effect then, the mass of a particle such as an electron is altered by the 
potential energy of the field. At the center of rotation (where the 
potential is zero) this mass is simply the usual one we are familiar with. 
Deep in a strong potential well, however, it will be markedly less. (Note 
that kinetic energy is positive and increases the mass, whereas potential 
energy is negative here and decreases the mass.) The mass change therefore 
will be

\begin{equation}
\label{eq4}
\Delta m = \frac{e}{c^2} - m \cong - \textstyle{1 \over 2}m\frac{\omega 
^2r^2}{c^2}.
\end{equation}

For $\omega r$ on the order of $c$, this is a significant, finite, and readily 
measurable change, even when $\omega $ and $\omega ^{2}r$ are 
infinitesimally small. We just measure the mass of an electron in an 
experiment and it will tell us the potential. That, in turn, tells us the 
circumferential speed of our rotating frame, and the degree of 
non-time-orthogonality.

This parallels the Schwarzchild field situation with its inherent time 
dilation and meter stick contraction, which are not dependent on the 
gravitational force felt, but on the potential. One could have a 
gravitational system with no detectable gravitational force, but significant 
gravitational potential. Hence, the effect on meter sticks and clocks would 
be significant and finite, even for an observer who could not discern any 
gravitational force field.\cite{One:1}

\subsection{NTO Answers to the Challenges}

The numbers below correspond to those of the challenges listed in section 
\ref{sec:challenges}.

1) My first reply to the non-limit case where a 
meter stick is suddenly released is that this is little different from the 
sudden change in stress in an object attached to a centrifuge that is 
suddenly released. Stress goes from being extremely high to zero in a 
theoretically infinitesimal time interval. (Actually, in the real world the 
time interval is more on the order of perhaps microseconds or so.) Why 
should Lorentz contraction be any different?

My second reply is that a similar situation occurs in the traditional 
approach. From earlier arguments in section \ref{sec:mylabel1}, it 
seems inescapable that the only feasible way Lorentz contraction could occur 
on a rotating disk is if it were absolute. That is, both lab and disk 
observers agree the disk meter sticks are shorter than the lab meter sticks. 
Then consider what the disk observer sees upon release. The lab meter sticks 
must look longer to him than his meter sticks when he is attached to the 
disk. But upon release, special relativity applies, and the lab meter sticks 
must then look shorter to him. This change from longer to shorter must, by 
the same logic, be instantaneous, and the traditional approach is therefore 
subject to identical presumed shortcomings as the NTO approach was 
considered to have.

My third reply to this challenge is that the transformation from a 
non-contracted meter stick (as in NTO theory) to a contracted one upon 
sudden release is not actually instantaneous. As explained in the Appendix, 
the difference in simultaneity between frames results in a gradual (relative 
to the speed of light) contraction of the meter stick beginning from the 
leading edge and propagating toward the trailing edge.

2) Regarding the limit case and the lack of ability 
to discern $\omega $ or $\omega ^{2}r$ because they are so small, I refer 
to the argument above in section \ref{subsec:additional}. One can still 
measure potential, hence $\omega r$, and therefore the degree of 
non-time-orthogonality. One then knows the frame in question is an NTO frame 
by measuring finite quantities. If the lab observer can look at the disk 
meter stick as it passes by, he can also look at a scale with an election on 
it (or Avagadro's number of some known type of atom) as it passes by. No 
inconsistency.

3)  To answer the combination of challenges 
1) and 2) we merely combine 
the answers to them above. We distinguish between pre and post release 
frames for the meter stick via measurements of the mass of a known particle 
type. That change is not infinitesimal and hence neither will be the rapid 
change as seen in the lab from uncontracted to Lorentz contracted state for 
the meter stick.

\section{Other Considerations}

In my discussions about NTO analysis with other physicists certain other, 
related, issues have been raised that I address below.

\subsection{Instantaneous Center of Rotation}

An object moving with variable velocity along a non-geodesic may twist and 
turn to various degrees along its path. At any given instant, however, one 
could measure its instantaneous values of $\omega $ and acceleration 
\textbf{a.} Taking $a=\omega ^{2}r$, one can then calculate an 
instantaneous value for \textbf{r}, the radial vector from the instantaneous 
center of rotation.

Note this value has physical significance in the sense Einstein considered 
so important in his gedanken elevator experiments. What one can measure 
solely inside one's frame, without looking outside, is central to the entire 
theory of relativity. This is the basis of the equivalence principle in 
which acceleration and gravitational force, being indistinguishable 
(locally), are equivalent.

Hence, from what we can measure inside the frame of our particle with 
variable velocity, we actually \textit{are} rotating about a center of rotation $r$ 
distance away\cite{Again:1}. Of course, if acceleration \textbf{a} changes an instant 
later, then so does \textbf{r}. In general, we can imagine \textbf{r} (and 
$\omega )$ continually ``jitterbugging'' all over the place. 

In our limit case example, upon release, the \textbf{r} vector for the meter 
stick would suddenly jump from extremely large to zero. One assumes here 
that the meter stick continues to rotate at $\omega $ after release.

More generally, an object undergoing general motion can be analyzed using 
NTO theory by determining the instantaneous center of rotation at each point 
in time during its trajectory, and applying the methodology of the theory at 
that instant. This then is a local theory, whose values may be integrated to 
determine globally measurable quantities.

Consider further, an object moving through space which in actuality would be 
continually impacting tiny dust and other type particles. Each impact would 
accelerate it and change its angular velocity. Hence, we could imagine a 
``lab'' observer watching this object. What would she see? Would the object 
look Lorentz contracted? Would it suddenly switch back and forth from 
Lorentz contracted to (according to NTO theory) not contracted? Would it 
never look contracted since to some (minute) degree it is always rotating 
and accelerating in some manner?

Note that in order for the velocity \textbf{v} we see in the lab for the 
object to be purely due to rotation the acceleration and angular velocity of 
the object must yield precisely \textbf{v}. To see the impact of this, 
consider a specific example, where an object is moving by us at $c$/3 and has

\begin{equation}
\label{eq5}
\omega = 10^{ - 20}\;{\rm rad / sec}\quad \quad \quad \omega ^2r = 10^{ - 
12}\;{\rm m / sec}^{\rm 2}
\end{equation}

\noindent
associated with it. Then

\begin{equation}
\label{eq6}
r = \frac{10^{ - 12}}{10^{ - 40}} = 10^{28}\,\,{\rm m}\quad \to \quad \omega 
r = 10^8\,\,{\rm m / sec} = \frac{c}{3},
\end{equation}

\noindent
and 100{\%} of the velocity we see for the object is due to rotation.

If the acceleration has another value that is still immeasurable, say one 
tenth of that above, then

\begin{equation}
\label{eq7}
r = \frac{\omega ^2r}{\omega ^2} = \frac{10^{ - 13}}{10^{ - 40}} = 
10^{27}\,\,{\rm m}\quad \to \quad \omega r = 10^7\,\,{\rm m / sec} = 
\frac{c}{30}
\end{equation}

\noindent
and only 10{\%} of the speed $c/$3 that we see for the object is due to its 
instantaneous rotation. Hence, we would see a Lorentz contraction as if the 
object had a translational speed of 90{\%} of the speed we see with respect 
to us.

The point is, that for all practical purposes the angular velocity and 
acceleration values, when very small, will virtually never combine to 
produce a rotational tangent velocity anywhere close to the velocity we see 
for the object passing by us. The velocity we see of any object moving 
through space where its rotation rate and acceleration are negligible can 
therefore be considered effectively due entirely to translation and all of 
the usual (time orthogonal) theory of relativity applies.

\subsection{Meter Stick Lengths in Three Frames}
\label{subsec:meter}

There is often confusion regarding the lengths of meter sticks seen by three 
different relevant observers: 1) the lab observer, 2) the rotating disk 
observer fixed to the rim, and 3) an inertial observer instantaneously 
co-moving with the disk rim. Each is carrying his/her own meter stick and 
each is looking at the meter sticks of the two others. What does each see?

We answer this question employing the graphical construction used in the 
Appendix. As noted there, this is well known means for illustrating the 
Lorentz contraction resulting from the Lorentz transformation. Here we 
extend it in Figure 1 below to NTO frames as well. For a derivation of this 
from the appropriate transformations, see Born \cite{Max:1965} and 
reference \cite{Robert:1998}.

We use the notation of ref \cite{Robert:1998}, where upper case 
represents inertial frames and lower case the rotating frame. K is the lab 
frame; K$_{1}$ that of an inertial observer (in a jet plane with engine off 
for illustration) instantaneously at rest with respect to a point on the 
disk rim; and k that of the observer on the disk rim point.

\begin{figure} [tbp]
\includegraphics[bbllx=-.74in,bblly=.13in,bburx=4.95in,bbury=4.80in,scale=1.00]{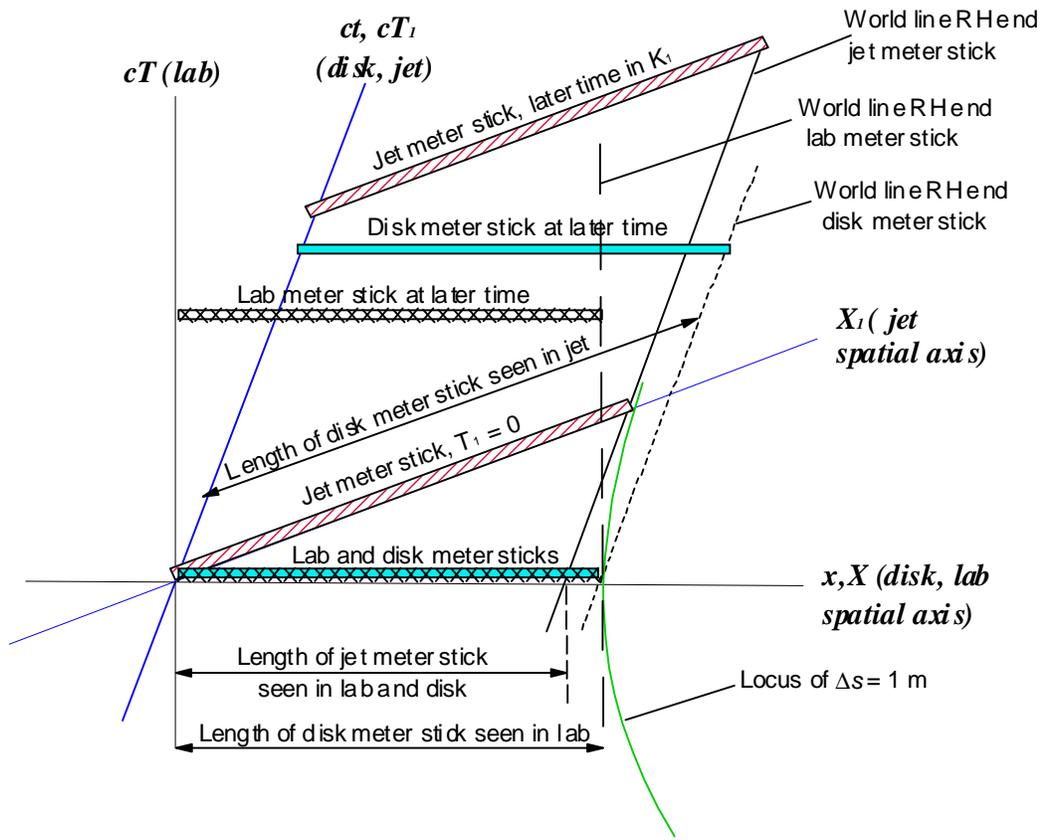}
\caption{. Length of Meter Sticks in Other Frames Seen by Given Observers}
\label{fig1}
\end{figure}

\bigskip

Note that the ends of each meter stick trace out world lines in spacetime. A 
given meter stick is considered ``seen'' by a particular observer when the 
endpoint events of that meter stick are simultaneous (occur at the same 
time) for the particular observer. Time intervals are considered 
infinitesimal such that the disk meter stick moves in essentially the same 
direction as the jet meter stick for the interval considered. From 
inspection of Figure 1, one can glean the significance with regard to 
Lorentz contraction of the NTO nature of the rotating frame.

\bigskip

The message of Figure 1 is summarized in Table 1 below. The left hand column 
of the table represents the observer. Each successive column represents the 
respective meter stick at rest in the given frame (top box). The comment in 
each box describes how the observer in her frame sees the meter stick length 
in the other given frame compared to the meter stick she is carrying with 
her in her own frame.

\bigskip

\begin{longtable}[htbp]
{|p{135pt}|p{99pt}|p{99pt}|p{94pt}|}
a & a & a & a  \kill
\hline
Observer below sees  \par meter sticks at right as& 
K (lab)& 
K$_{1}$ (jet)& 
k (disk) \\
\hline
K (lab)& 
--& 
contracted& 
same \\
\hline
K$_{1}$ (jet)& 
contracted& 
--& 
longer \quad  \\
\hline
k (disk)& 
same& 
contracted& 
-- \\
\hline
\caption{Length of Meter Sticks in Other Frames\\               Seen by Given Observers}
\label{tab1}
\end{longtable}

\bigskip

Also, by comparison of columns above, we see the following.

The lab observer in K sees the K$_{1}$ (jet) meter stick as shorter than the 
k (disk) meter stick.

The jet observer in K$_{1}$ sees the K (lab) meter stick as much shorter 
than the k (disk) meter stick.

The disk observer in k sees the K$_{1}$ (jet) meter stick as shorter than 
the K (lab) stick.

\section{Summary and Conclusions}

From the logic presented herein, it appears that the traditional approach 
prediction of Lorentz contraction on the rim of a rotating disk is 
inconsistent. The NTO approach prediction of no Lorentz contraction appears 
consistent. These conclusions apply to both the limit and non-limit cases.

The rotating frame limit case has been demonstrated distinguishable from a 
Lorentz frame due to substantial and measurable decrease in mass of known 
particle types, which results from the high value for the potential at large 
radius.

General motion of particles may be described using traditional local Lorentz 
frame methodology in conjunction with NTO analysis. For NTO analysis, the 
instantaneous center of rotation is determined from the angular velocity 
$\omega $ and the particle acceleration \textbf{a}. The degree of 
non-time-orthogonality is then found from $\omega r$.

The effectively physical local metric (associated with coordinates whose 
values are those one would measure via experiment with physical instruments, 
i.e., standard meter sticks and clocks) has a general form with unit values 
for diagonal components and non-zero off diagonal spacetime components. In 
the special (but far more commonplace) case when time is orthogonal to 
space, this metric reduces to the Minkowski (or Lorentz) metric.

\section*{Appendix: Transitional Release of a Rotating Frame Meter Stick}

Figure 2 depicts the sudden release of a meter stick from a rotating frame 
as seen by an observer in the lab frame. The graphical construction used is 
well known and is described briefly in Section \ref{subsec:meter} 
as well as more extensively in references cited therein.

The release is considered to happen at time $T$ = 0 in the lab frame. The meter 
stick is shown as observed at a time prior to $T$ = 0 in the portion of the 
graph below the horizontal (spatial) axis. It is shown again above that as 
observed just as it is released (endpoint events O and N).

A meter stick traveling in the Lorentz frame having velocity equal to the 
instantaneous tangent velocity $\omega r$ would be observed to have end 
events O and P for an observer in that frame. As the meter stick is released 
it changes from the rotating frame state to the moving (relative to us, the 
observers) Lorentz frame state.

\begin{figure} [tbp]
\includegraphics[bbllx=-.74in,bblly=0.13in,bburx=4.25in,bbury=3.92in,scale=1.00]{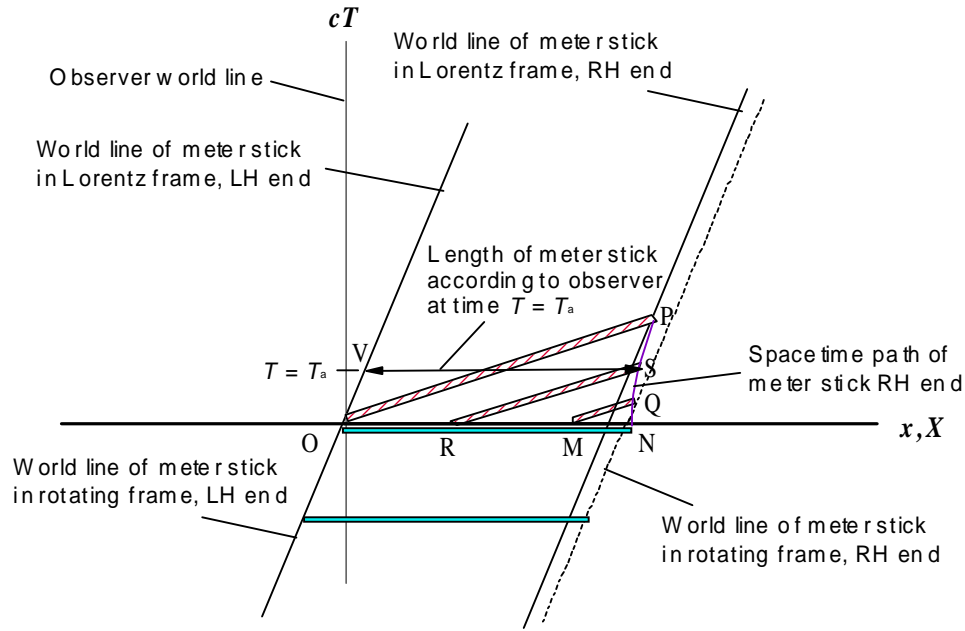}
\caption{. Meter Stick Suddenly Released from Rotating Frame}
\label{fig2}
\end{figure}

\bigskip

The release of the meter stick endpoints comes at events O and N. Note that 
within the moving Lorentz frame (the ``jet'' frame of Section 
\ref{subsec:meter}) these events are not simultaneous. N occurs 
before O therein. Event M occurs shortly after N for the jet frame, but is 
simultaneous with N for the lab. Using the standard construction from event 
M, we obtain the representation MQ, where MQ has the same 4D length as MN. 
Similarly, RS has the same spacetime length as RN; and OP has the same 
spacetime length as ON. Note that at time $T_{a}$, the length for the meter 
stick seen in the lab would be VS, and this is part way between 
non-contracted (as for the disk frame) and Lorentz contracted (as for the 
jet frame.) The result is a smooth transition of meter stick length as 
observed in the lab from uncontracted to Lorentz contracted.


\begin{thebibliography}{21}
\bibitem{Relativistic:1977} Gr{\o}n, {\O}., "Relativistic Description of a Rotating Disk", \textit{Am. J. Phys.} 43(10), 869-876 (1975); "Rotating Frames in Special Relativity Analyzed in Light of a Recent Article by M. Strauss"\textit{,} \textit{Int'l Journal of Theoretical Physics }\textbf{16}(8), 603-614 (1977).
\bibitem{Einstein:1938} Einstein, A., and Infeld, L., \textit{The Evolution of Physics} (Simon and Schuster, 1938), pp. 226-234.
\bibitem{Stachel:1980} Stachel, J., "Einstein and the Rigidly Rotating Disk"\textit{,} Chapter 1 in Held, \textit{General Relativity and Gravitation} (Plenum Press, New York, 1980), pp. 1-15.
\bibitem{Weber:1997} T.A. Weber, ``Measurements on a rotating frame in relativity, and the Wilson and Wilson experiment'', \textit{Am. J. Phys.} 65 (10), 946-953 October 1997.
\bibitem{Pathra:1} Pathra, \textit{The Theory of Relativity}, eq (6.4), p. 156.
\bibitem{Richard:1944} Richard A. Mould, \textit{Basic Relativity}, Springer-Verlag, N.Y. (1944), p. 272.
\bibitem{Actually:1} Actually, as speed increases, the disk rim presumably tries to contract, but is prevented from doing so by the structural integrity of the disk material. In this view stress is thus induced circumferentially in the disk purely from relativistic kinematic causes. Increasing speed increases stress and at a certain angular velocity the disk should rupture. However, independent rods or meter sticks laid down around the circumference that are not connected to one another should each Lorentz contract independently, directly, and unfettered. In either case, we shall simplify our discussion by referring to this phenomenon as simply Lorentz contraction of the rim.
\bibitem{Arzeli:1966} Arzeli\`{e}s, H., \textit{Relativistic Kinematics,} (Pergamon Press, New York 1966), Chapter IX. Arzeli\`{e}s was not convinced that tension arises in the disk, but references others who were.
\bibitem{Paul:1937} Paul Langevin, ``Sur la th\'{e}orie de relativit\'{e} et l'experience de M. Sagnac'',\textit{ Academie des sciences comptes rendus des seances., }Vol. 173, 831-834 (7 Nov 1921); ``Relativit\'{e} -- Sur l'experience de Sagnac'',\textit{ Academie des sciences comptes rendus des seances., }Vol. 205, 304-306 (2 Aug 1937.)
\bibitem{Robert:1998} Robert D. Klauber, ``New perspectives on the relatively rotating disk and non-time-orthogonal reference frames'', \textit{Found. Phys. Lett. }\textbf{11}(\ref{eq5}) 405-443 (1998). qr-qc/0103076
\bibitem{Neil:1997} Neil Ashby, ``Relativity and the Global Positioning System'', \textit{Phys. Today}, May 2002, 41-47. See pg 44; , ``Relativistic Effects in the Global Positioning System'', \textit{15}$^{th}$\textit{ Intl. Conf. Gen. Rel. and Gravitation,} Pune, India (Dec 15-21, 1997), available at www.colorado.edu/engineering/GPS/Papers/RelativityinGPS.ps. See pp. 5-7. Ashby observes that applying the assumptions implicit in the traditional approach to the rotating earth frame produce errors in the GPS. Use of the NTO approach, however, yields correct results.
\bibitem{Brillet:1979} A. Brillet and J. L. Hall, ``Improved laser test of the isotropy of space,'' \textit{Phys. Rev. Lett}., \textbf{42}(9), 549-552 (1979). Brillet and Hall report a persistent signal they designate as ``spurious'' because it is not predicted by the traditional approach. However, such a signal is predicted by the NTO approach.
\bibitem{Sagnac:1914} M. G. Sagnac, \textit{Comptes Rendus, }\textbf{157}, 708-718 (1913); ``Effet Tourbillonnaire Optique. La Circulation de l'\'{e} lumineaux dans un Interf\'{e}rographe Tournant'', \textit{Journal de Physique Th\`{e}orique et Appliq\`{e}e}, Paris, Soci\`{e}te fran\c{c}aise de physique, Series 5, Vol 4 (1914), 177-195. Sagnac states clearly and repeatedly that from the point of view of the rotating apparatus he considers his experiment inexplicable via the traditional approach. 
\bibitem{Alexandre:1942} Alexandre Dufour et Fernand Prunier, ``Sur l'observation du ph\'{e}nom\'{e}ne de Sagnac avec une source \'{e}clairante non entra\'{\i}n\'{e}e'', \textit{Academie des sciences comptes rendus des seances}., Vol. 204, 1322-1324 (3 May 1937); ``Sur un Deplacement de Franges Enregistre sur une Plate-forme en Rotation Uniforme'', \textit{Le Journal de Physique et Le Radium}, serie VIII, T. III, No 9, 153-161 (Sept 1942). Dufour and Prunier repeat Sagnac's experiment under a variety of conditions. They too are emphatic in their belief that the results can not be predicted via the traditional approach.
\bibitem{Robert:1} Robert D. Klauber, ``Derivation of the General Case Sagnac Experimental Result from the Rotating Frame'', gr-qc/0206033. The Sagnac experimental result is derived using the NTO approach. It is submitted that this result has never been derived from the rotating frame using the traditional approach, and that it does not appear possible to do so. Langevin (ref \cite{Paul:1937}) drew similar conclusions.
\bibitem{Ref:1} Ref \cite{Robert:1998} summarizes the similarities of, and differences between, the predictions of the traditional and NTO approaches.
\bibitem{Tartaglia:1999} Tartaglia, A., ``Lengths on Rotating Platforms'', \textit{Founds. Phys. Lett., }\textbf{12}(\ref{eq1}), 17-28 (1999). Tartaglia and I have made the point in this paragraph independently.
\bibitem{NTO:1} NTO analysis resolves this issue by limiting the validity of the surrogate local Lorentz frames approach to (non-inertial or inertial) frames in which time is orthogonal to space. While for NTO frames this heretofore sacrosanct principle appears invalid, it continues to be correct for the vast majority of all cases treated in general relativity.
\bibitem{For:1} For example, due to time dilation, the angular velocity $\omega _{k}$ measured inside the rotating frame k would be greater than the angular velocity $\omega $ as seen from the non-rotating frame. In particular, $\omega _k = \omega / \sqrt {1 - \omega ^2r^2 / c^2} $. Since relativistically, it can be shown that $a_k = \omega _k^2 r$, one can solve these two equations for \textit{r} and $\omega $, the values typically employed in NTO analysis.
\bibitem{NTO:2} NTO theory is based on differential geometry, just as is traditional time orthogonal general relativity. The same result is obtained in both. See ref. \cite{Robert:1998}, section 4.3.4, eq (26).
\bibitem{One:1} One might consider that the change in mass measured might be due to a gravitational potential rather than a rotational potential, and thus the system would not be subject to the idiosyncracies of NTO systems. That is, the decrease in mass would not have to be due to a rotational potential \textit{per se}. However, in a gravitational potential \textit{both} observers are immersed in the gravitational field. Hence, each would measure the same mass change for a given particle type at rest with respect to her. For rotation, only the disk based observer would measure such a change. Looking at a mass measurement experiment in the other's frame and comparing with the same experiment in one's own frame would readily determine which observer is truly in a rotating frame.
\bibitem{Again:1} Again, we are illustrating a principle using relations from non-relativistic mechanics. The same principle is readily shown to be true relativistically.
\bibitem{Max:1965}Max Born, \textit{Einstein's Theory of Relativity}, Dover, NY, 238-255 (1965).
\end{thebibliography}
\end{document}